\documentclass[
aps,
prb,
reprint,
superscriptaddress,
showkeys,
amsmath, 
amssymb,  
longbibliography, 
nofootinbib,
]{revtex4-1}
\usepackage{dcolumn}
\usepackage[toc,page]{appendix}
\usepackage{graphicx}
\usepackage[bottom]{footmisc}
\usepackage{dcolumn}
\usepackage{bm}
\usepackage{comment}
\usepackage{booktabs}
\usepackage{array}
\usepackage{multirow}
\usepackage{makecell}
\usepackage{amsmath}
\usepackage{indentfirst}
\usepackage{hyperref}
\usepackage{float}

\begin{document}

\preprint{APS/123-QED}

\author{Samantha I. Davis}
\thanks {These authors contributed equally to this work}
\author{Rahim R. Ullah}
\thanks {These authors contributed equally to this work}
\altaffiliation[current address: ]{Department of Physics, University of California, Davis, 1 Shields Ave, Davis, CA 95616, USA}
\affiliation{Department of Physics, Stanford University, Stanford, CA 94305, USA}

\author{Carolina Adamo}
\affiliation{Geballe Laboratory for Advanced Materials, Stanford University, Stanford, CA 94305, USA}
\author{Christopher A. Watson}
\affiliation{Department of Applied Physics, Stanford University, Stanford, CA 94305, USA}

\affiliation{Stanford Institute for Materials and Energy Sciences, SLAC National Accelerator Laboratory, 2575 Sand Hill Road, Menlo Park, CA 94025, USA}
\author{John R. Kirtley}
\affiliation{Geballe Laboratory for Advanced Materials, Stanford University, Stanford, CA 94305, USA}
\author{Malcolm R. Beasley}
\affiliation{Geballe Laboratory for Advanced Materials, Stanford University, Stanford, CA 94305, USA}
\author{Steven  A. Kivelson}
\affiliation{Department of Physics, Stanford University, Stanford, CA 94305, USA}
\affiliation{Stanford Institute for Materials and Energy Sciences, SLAC National Accelerator Laboratory, 2575 Sand Hill Road, Menlo Park, CA 94025, USA}
\author{Kathryn A. Moler}
\affiliation{Department of Physics, Stanford University, Stanford, CA 94305, USA}
\affiliation{Geballe Laboratory for Advanced Materials, Stanford University, Stanford, CA 94305, USA}
\affiliation{Department of Applied Physics, Stanford University, Stanford, CA 94305, USA}
\affiliation{Stanford Institute for Materials and Energy Sciences, SLAC National Accelerator Laboratory, 2575 Sand Hill Road, Menlo Park, CA 94025, USA}

\title{Spatially modulated susceptibility in thin film La$_{2-x}$Ba$_x$CuO$_4$}

\date{\today}

\begin{abstract}

The high critical temperature superconductor Lanthanum Barium Copper Oxide (La$_{2-x}$Ba$_x$CuO$_4$ or LBCO) 
exhibits a strong anomaly in critical temperature at 1/8th doping, nematicity, and other interesting properties. We report here Scanning Superconducting Quantum Interference Device (SQUID) imaging of the magnetic fields and susceptibility in a number of thin film LBCO samples with doping in the vicinity of the 1/8th anomaly.  Spatially resolved measurements of the critical temperatures of these samples do not show a pronounced depression 
at 1/8th doping. They do, however, exhibit strong, nearly linear modulations of the susceptibility (``striae'') of multiple samples with surprisingly long periods of $1-4~\mu$m. 
Counterintuitively, vortices trap in positions of largest diamagnetic susceptibility in these striae. Given the rich interplay of different orders in this material system and its known sensitivity to epitaxial strain, we propose phase separation as a possible origin of these features and discuss scenarios in which that might arise.

\end{abstract}

\pacs{Valid PACS appear here}
\maketitle
%

\section{\label{sec:level1}Introduction}


Since the discovery of high temperature superconductivity in Lanthanum Barium Copper Oxide compounds (LBCO) in 1986,\cite{BednorzPRB86} the cuprate perovskites and other unconventional superconductors have attracted enormous interest, not only because of their technological promise but also as a laboratory for exploring concepts in condensed matter physics. Although much progress has been made in understanding high temperature superconductivity,\cite{Tsuei2000,norman2003electronic,lee2007high,LeeJSNM2017} further development demands the empirical exploration of the properties of these materials.

Here, we have probed the phase diagram of thin film La$_{2-x}$Ba$_{x}$CuO$_{4}$ over a nominal doping range of $x_{nom}=0.090$ to $x_{nom}=0.135$. 
La$_{2-x}$Ba$_{x}$CuO$_{4}$ is a system of unconventional superconductors  that exhibit $d$-wave superconductivity\cite{Tsuei2000} at temperatures up to 35 K.\cite{BednorzPRB86}  Along with other unconventional properties, this renders the standard Bardeen-Cooper-Schrieffer theory insufficient to explain the underlying physics and emergent phenomena of such materials. In particular, a proper account of phenomena such as nematicity,\cite{nie2014quenched} pair density wave “stripes”,\cite{Berg2009charge} and the one-eighth anomaly\cite{fujita2002competition} pose challenges to current theories of superconductivity. 
Here we studied six samples of thin film LBCO with thicknesses of approximately 20~nm (see Table \ref{table:fits}), which were grown on a nearly lattice matched substrate. We used scanning SQUID microscopy\cite{kirtley2010fundamental} to measure the magnetic fields and susceptibilities of the samples over a range of temperatures. We observed striking oscillations (``striae") in the superconducting susceptibility and correlated the behaviors of these striae with other properties of the films.

\section{Experimental Methods}
\begin{figure}[b]
\includegraphics[width=\columnwidth]{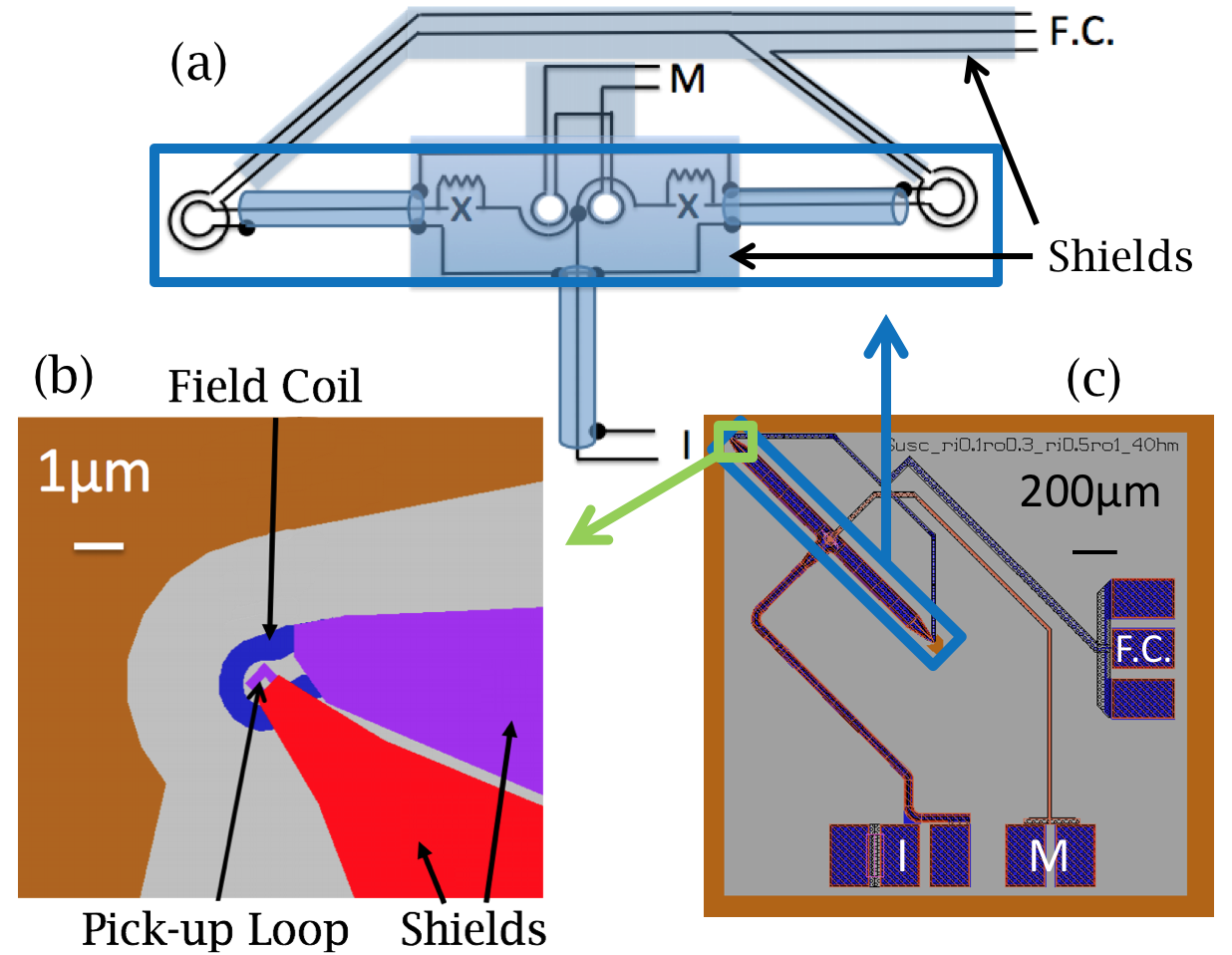}
\caption{{\bf SQUID Susceptometer with 0.1 $\mu$m radius pickup loop}\cite{kirtley2012scanning}(a) Schematic layout. The current leads, modulation coil, and field coil are labeled by $I$, $M$, and $F.C.$ respectively. All but the pickup loop/field coil regions are shielded from external magnetic fields by superconducting layers. (b) Layout of the pickup loop/field coil region. (c) Layout of the entire 2 mm $\times$ 2 mm chip.}
\label{fig:squidModel}
\centering
\end{figure}

\begin{figure}[h]
\centering
\includegraphics[width=\columnwidth]{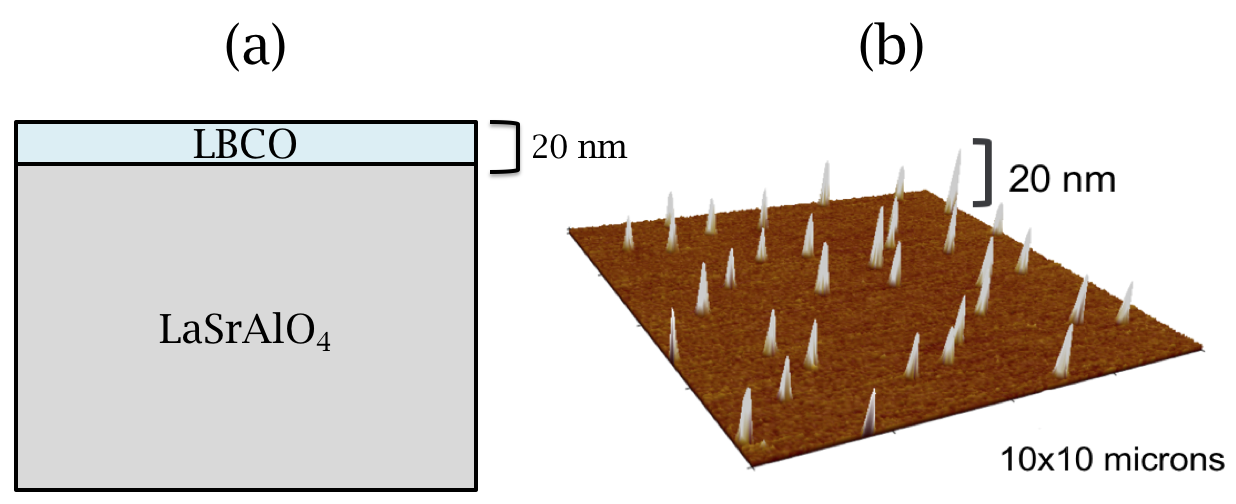}
\caption{{\bf Sample geometry} (a) Schematic of 20~nm thick LBCO film on a LaSrAlO$_4$ substrate. (b) AFM image of AC 174, a 20~nm thick film of La$_x$Ba$_{2-x}$CuO$_4$ with nominal doping of x$_{\rm nom}$~=~0.125.}
\label{fig:AFM}
\end{figure}

Our microscope implements a D.C. SQUID
with small pickup loops integrated into the body of the SQUID through well-shielded coaxial superconducting leads, allowing for high spatial resolution magnetometry measurements.\cite{kirtley1995high} 
The addition of a field coil, which is geometrically arranged on the SQUID chip such that it applies a controlled amount of magnetic flux to the sample but zero net flux to the SQUID, allows us to also simultaneously measure the sample\textsc{\char13}s magnetic response to an applied magnetic field – i.e., its magnetic susceptibility.\cite{gardner2001scanning}

To explore the local superconducting properties of thin film LBCO, we used a SQUID with a 0.1 $\mu$m inner and 0.3 $\mu$m outer pickup loop radius, and inner and outer field coil radii of 0.5$\mu$m and 1.0$\mu$m respectively.\cite{kirtley2016scanning} This enabled us to spatially resolve the diamagnetism of our samples with submicron spatial resolution, 
thus resolving the micron-scale oscillations in the susceptibility of our samples. 

The thin films of La$_{2-x}$Ba$_{x}$CuO$_{4}$ were grown using shuttered layer-by-layer\cite{haeni2000RHEED} deposition on (100) LaSrAlO$_4$ single crystal substrates in a reactive GENxplor VEECO molecular-beam epitaxy system. A substrate temperature of 750$^\circ$ C and an oxygen plasma background partial pressure of $2\times 10^{-6}$ Torr, which was kept constant until the temperature of the substrate dropped below 150$^\circ$C, were used.
Ba was evaporated using a low temperature effusion cell, and Cu and La were evaporated using high temperature effusion cells. The flux of each element was measured by a quartz crystal monitor (QCM) before the growth, and these measurements were used to determine the nominal doping. Attempts to confirm the doping with XPS led to inconsistent values, possibly due to the difficulty of making XPS measurements on such thin films. Reflection high-energy electron diffraction (RHEED)\cite{haeni2000RHEED} was used to monitor both the phase purity and the stoichiometry during film growth.  RHEED oscillations taken during fabrication of the films indicate that the samples have a smooth surface. Atomic Force Microscopy (AFM) measurements (see Fig. \ref{fig:AFM}) reveal that the samples are flat except for 20 nm thick bumps dispersed over the surface. The measured samples are shown in Table 1.

We used a liquid helium scanning SQUID microscope system to image the magnetic fields and susceptibilities at different locations on our samples.\cite{attocube} This microscope allowed us to vary the sample temperature over a wide range while keeping the SQUID sensor superconducting.\cite{kirtley1999variable}

\begin{table}[ht]
\caption{{\bf Sample parameters} Listed are the  sample names, nominal dopings $x_{\rm nom}$, superconducting critical temperatures T$_c$, the effective field coil radius $R$ divided by the Pearl length $\Lambda$ measured at 5 K, and striae period.}
\centering 
\renewcommand{\arraystretch}{1.5}
\begin{tabular}{| c | c | c | c | c |}

\hline
\textbf{Sample} & \textbf{$x_{\rm nom}$}& \textbf{T\boldmath$_c(K)$} &  \boldmath$R/\Lambda(5 K)$  & \textbf{ Period (\boldmath$\mu$m)}\\
 [0.5ex]
\hline
\midrule
AC 201 & 0.090 & 22.9$\pm$1.1   & 0.02$\pm$0.005   & 2.6-2.7\\
[1ex]
\hline
AC 202 & 0.098 & $<$5           & -                & -      \\
[1ex]
\midrule
\hline
AC 173 & 0.115 & 23.7$\pm$0.2   & 0.23$\pm$0.01    & 3.4-3.8\\
[1ex]
\midrule
\hline
AC 174 & 0.125 & 29.2$\pm$0.2   & 0.26$\pm$0.01   & 1.0-2.6\\
[1ex]
\hline
AC 200 & 0.125 & 29.7$\pm$1.4   & 0.02$\pm$0.005   &  -     \\
[1ex]
\hline
AC 175 & 0.135 & 27.8$\pm$0.3   & 0.17$\pm$0.01   &      1.2\\
[1ex] 
\hline 
\end{tabular}
\label{table:fits} 
\end{table}

\section{Results}
\begin{figure}
\centering
\includegraphics[width=3.5in]{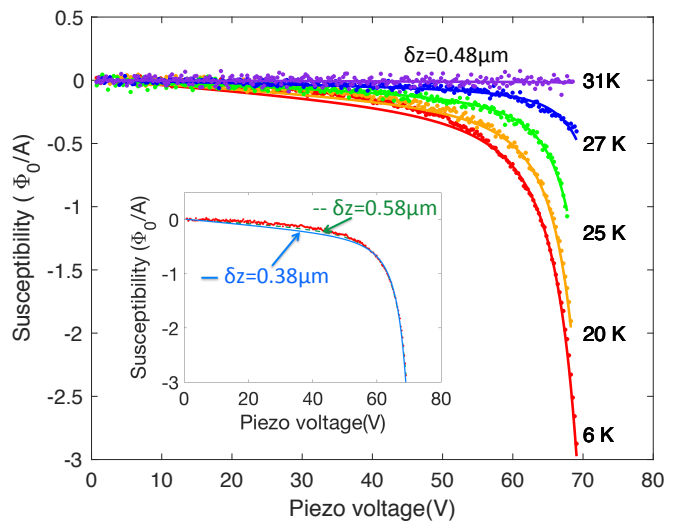}
\caption{{\bf Susceptibility touchdowns vs. temperature} Measurements of the change in the mutual inductance $\phi=\Phi/I$ (in units of $\Phi_0=h/2e$) between the field coil and the pickup loop in our SQUID susceptometers for sample AC 174 as a function of the z-piezo voltage. The sample comes into mechanical contact with the SQUID substrate at about 70 V (with a z-piezo scanner calibration constant of 9.4~V/$\mu$m). The dots are data, at temperatures as labeled. The solid lines are fits to Eq. \ref{eq:tdthindia}. For these fits we took the measured value $\phi_s = 54 \Phi_0$/A, the effective field coil radius $R$ = 0.79$\mu$m,\cite{kirtley2016scanning} the spacing $\delta z$ = 0.48~$\mu$m, with $\phi_{\rm offset}$, $\alpha$ and $R/\Lambda$ as fitting parameters. The inset replots the 6K data, with best fits for $\delta$z = 0.38$\mu$m (blue solid line) and 0.58$\mu$m (green dashed line). }
\label{fig:touchdowns}
\end{figure}

\begin{figure}
\centering
\includegraphics[width=3.5in]{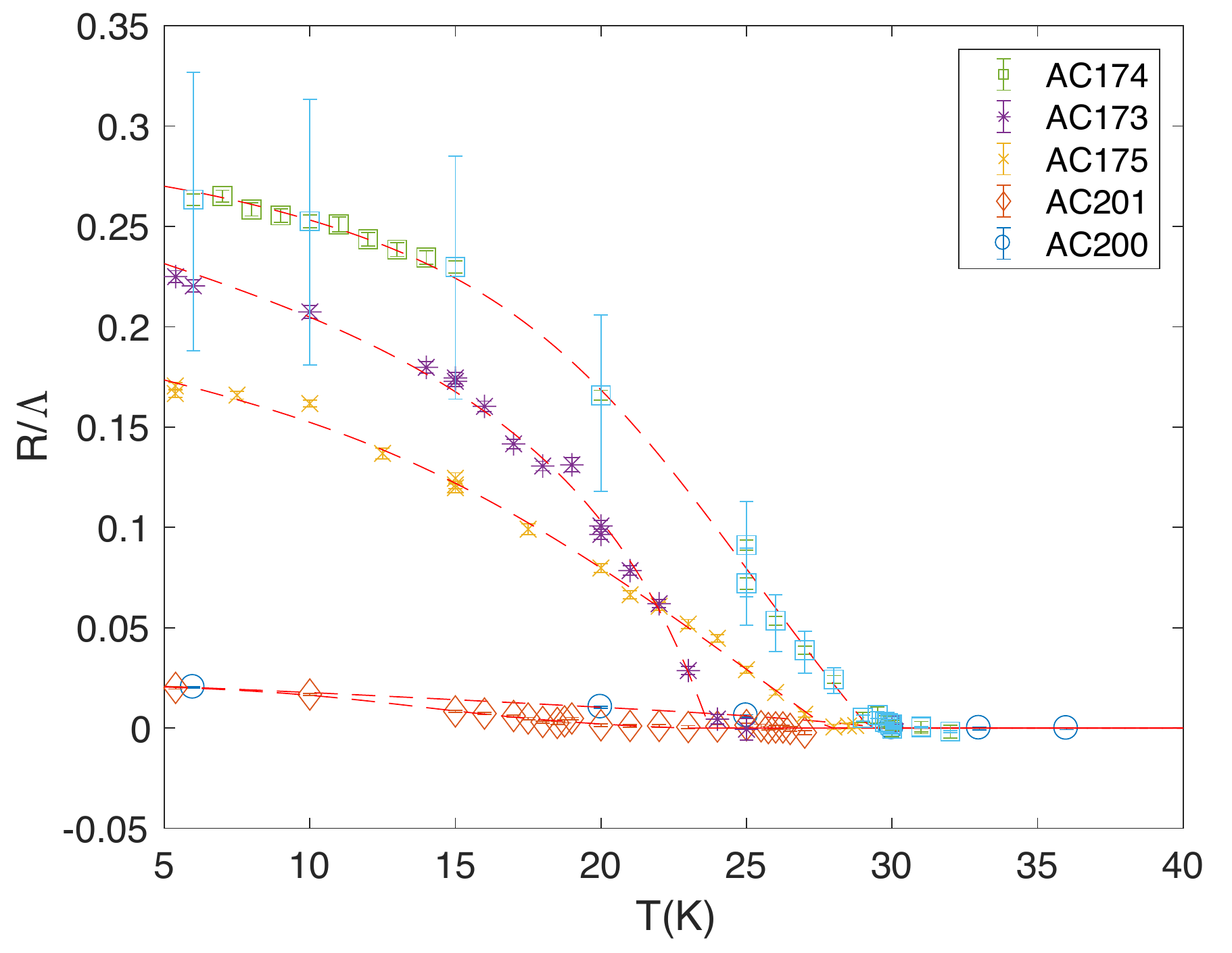}
\caption{{\bf Pearl length vs. temperature} The symbols are fit values for R/$\Lambda$ from touchdown data as illustrated in Fig. \ref{fig:touchdowns}, assuming the sample surface to pickup loop distance at contact $\delta z = 0.48 \mu m$. The dashed lines are guides for the eye. The error bars on the data for AC174 represent systematic errors due to uncertainty in $\delta z$, with the lower and upper bars corresponding to $\delta$z = 0.38 $\mu m$ and 0.58 $\mu m$ respectively.}
\label{fig:RLambdafits}
\end{figure}

\subsection{Pearl length}
We determined the local Pearl length and critical temperature of a sample by moving it towards the SQUID using a z-piezo scanner until contact was reached while measuring the SQUID susceptibility. An example for sample AC174 (see Table \ref{table:fits}) is shown in Fig. \ref{fig:touchdowns}. Here a 1 mA, 928 Hz A.C. current was passed through the field coil. The resultant in-phase flux signal $\phi=\Phi/I$, where $\Phi$ is the flux through the SQUID and $I$ is the current through the field coil,
was phase sensitively detected and plotted vs. the z-piezo scanner voltage as the dots in Fig. \ref{fig:touchdowns}. The solid lines in Fig. \ref{fig:touchdowns} are fits to the data using an expression appropriate for a homogeneous thin film diamagnet:\cite{kirtley2012scanning}
\begin{equation}
\phi=\phi_{\rm offset}+\alpha(z-z_0)-\phi_s(R/\Lambda)(1-2\bar{z}/\sqrt{1+4\bar{z}^2}),
\label{eq:tdthindia}
\end{equation}
where $\phi_{\rm offset}$ is a constant offset in the susceptibility, due to e.g. a mismatch in the mutual inductance between the two pairs of field coils/pickup loops, $\alpha$ is a coefficient of a small linear slope in the approach curves, $z_0$ is the position of the sample when it is in direct mechanical contact with the SQUID substrate, $\phi_s$ is the mutual inductance between one field coil and one pickup loop, $R$ is the effective radius of the field coil, $\Lambda$ is the Pearl length of the thin film, and $\bar{z}=(z_0+\delta_z-z)/R$, where $\delta_z$ is the physical spacing between the sample surface and the plane of the pickup loop when the sample and SQUID surface are in contact.\cite{kirtley2012scanning}  We estimate that $\delta z=0.48\pm0.1~\mu m$. The solid lines in the main panel in Fig. \ref{fig:touchdowns} are fits assuming $\delta z= 0.48~\mu m$. The inset of Fig. \ref{fig:touchdowns} shows how the fit quality changes when $\delta z$ is varied between 0.38 $\mu m$ and 0.58 $\mu m$.

The fitting parameters $R/\Lambda$ for the five samples which showed measurable diamagnetic shielding at 5 K are plotted as the symbols in Fig. \ref{fig:RLambdafits}. The dashed curves in Fig. \ref{fig:RLambdafits} are guides to the eye. Estimates for $R/\Lambda(T = 5~K)$ and $T_c$ are in Table \ref{table:fits}.
Since for a homogeneous thin film and in the absence of fluctuations the Pearl length $\Lambda=2\lambda^2/d$, where $\lambda$ is the London penetration depth and $d$ is the film thickness, and the superfluid density $n_s$ is given by $n_s = m/\mu_0 q^2 \lambda^2$, where $m$ is the mass and $q$ is the charge of the superconducting charge carriers, it follows that $n_s(0)/n_s(T) = \Lambda(T)/\Lambda(0)$,  and the fitting parameter $R/\Lambda(T)$ is a measure of the superfluid density $n_s(T)$.

An initially surprising result is that, although the low temperature Pearl length varies significantly from sample to sample, the measured critical temperatures of our samples do not, even though their nominal dopings span the $x_{\rm nom}=1/8$ range where a sharp drop in critical temperature is observed in bulk LBCO samples. Previous critical temperature measurements on thin film LBCO samples grown on LSAO similarly did not show a 1/8th doping anomaly.\cite{sato2000absence} 

\subsection{Striae in susceptibility}
\begin{figure*}
\begin{minipage}{\textwidth}
\includegraphics[width=\textwidth]{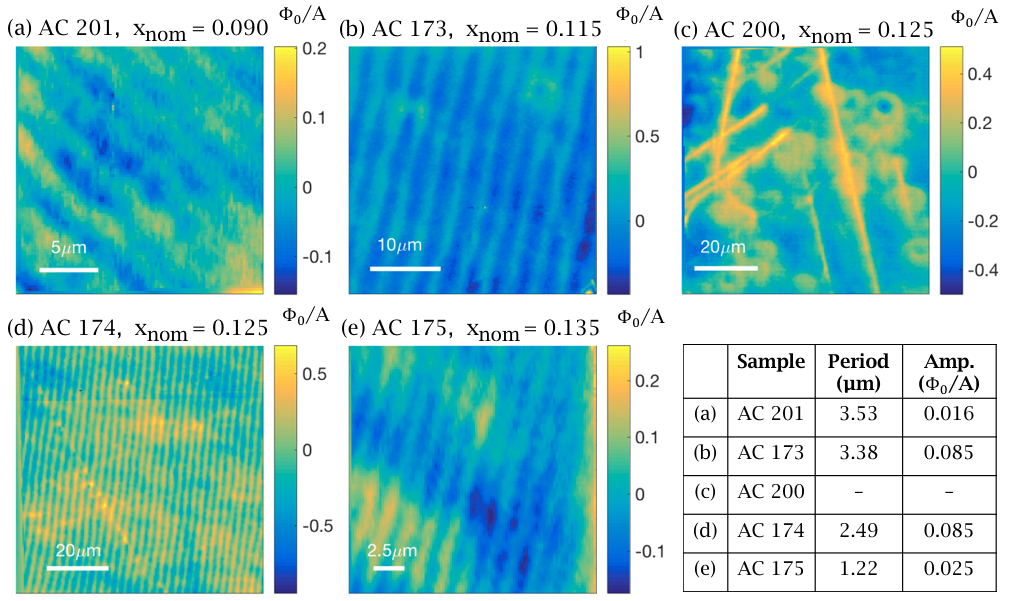}
\caption{ {\bf Striae are observed in four of the five superconducting samples imaged.} Magnetic susceptibility images of thin-film LBCO for distinct samples with nominal dopings of $0.090<x_{\rm nom}<0.135$. The ``striae'' are the periodic modulations in susceptibility shown in (a), (b), (d), and (e). Note the different scale bars: striae periods between one and four microns were observed in these samples. An average background susceptibility was subtracted out from each of these images. The amplitudes of the striae are approximately 1\%, 8\%, 4\%, and 1\% of the total susceptibility of scans (a), (b), (d), and (e), respectively. 
}
\label{fig:suscscanz}
\end{minipage}
\end{figure*}

\begin{figure}
\centering
\includegraphics[height=6in]{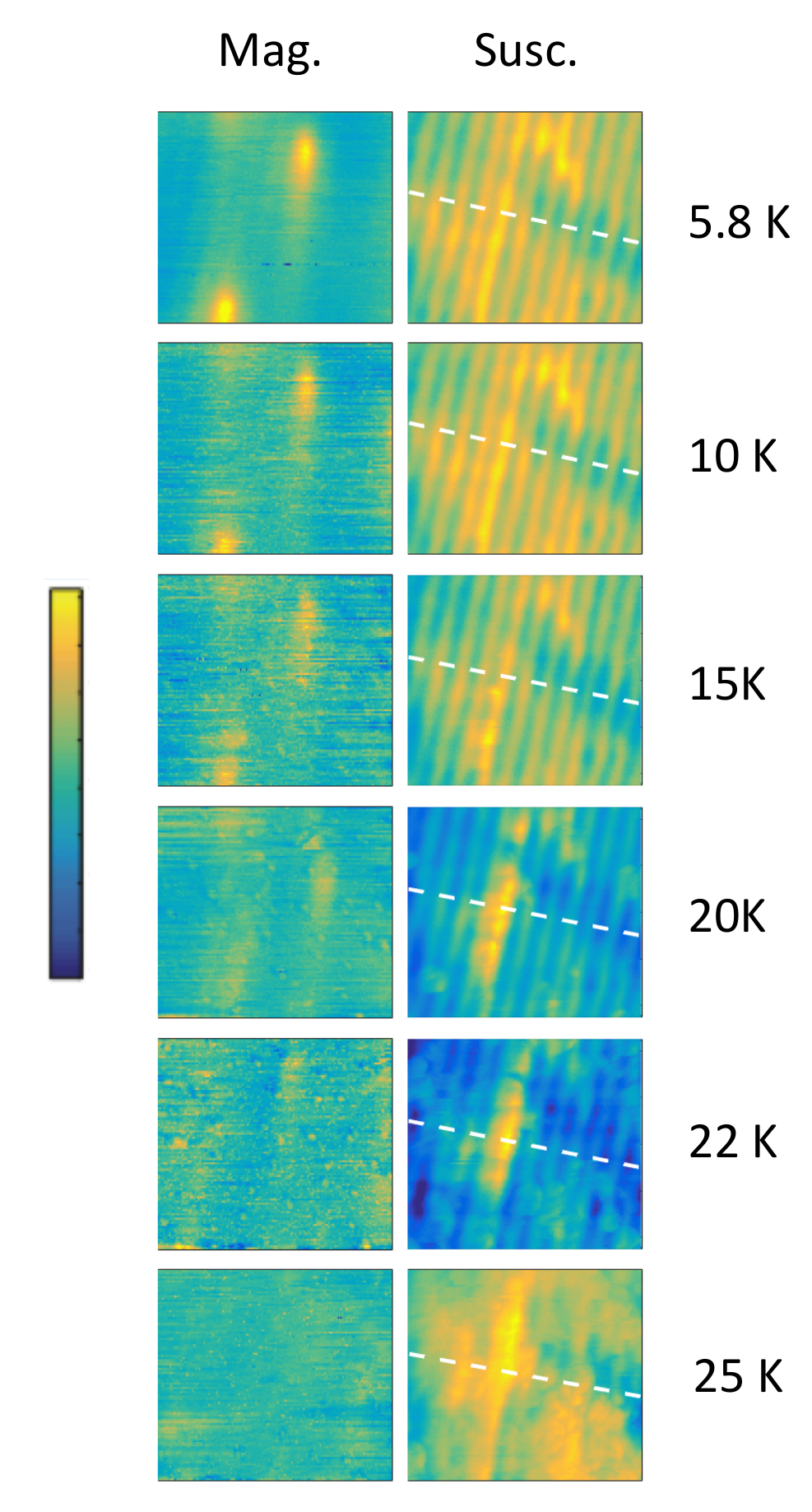}
\caption{{\bf Striae persist, with little change in period, to close to T$_c$} Magnetometry (left column) and susceptibility (right column) images of a 10~$\mu$m wide by 9~$\mu$m high area of sample AC174, cooled and imaged in a field of 33 $\mu$T at the temperatures labeled, with a field coil modulation of 10~mA at 2.204~kHz. The false color variations are 4~m$\Phi_0$ for the magnetometry images, and 0.4~$\Phi_0$/A for the susceptibility images. The white dashed lines represent the cross-sections through the data displayed in Fig.~\ref{fig:cross_vs_T}.}
\label{fig:mag_susc_vs_T}
\end{figure}

\begin{figure}
\centering
\includegraphics[width=3.50in]{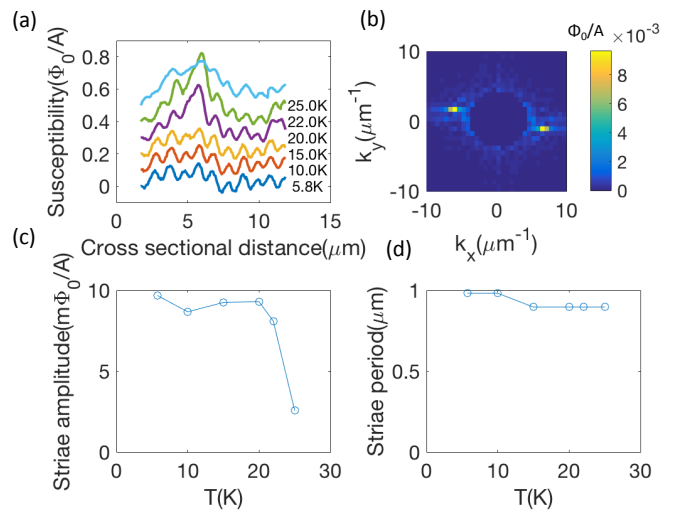}
\caption{{\bf Temperature dependence of the striae.} (a) Cross-sections through the susceptibility data of Fig. \ref{fig:mag_susc_vs_T} at the temperatures labeled. (b) two-dimensional Fourier transform of the 5.8 K data: the sharp peaks represent the susceptibility striae, with 1~$\mu$m period, amplitude 0.01~$\Phi_0$/A, and angle of -0.22 radians. (c) Plot of the stripe amplitude, from such Fourier transforms as in (b), as a function of temperature T. (d) Plot of the stripe period vs. T.}
\label{fig:cross_vs_T}
\end{figure}

Our susceptibility imaging results are also surprising: we find periodic “striae”  of modulated diamagnetism in the susceptibility images of multiple samples. Examples for the five samples with observable diamagnetic susceptibility at 4K are displayed in Fig. \ref{fig:suscscanz}. 
These striae are modulations of the SQUID susceptibility with amplitudes of $1-8\%$ of the total susceptibility and periods from $1-4~\mu$m. In the four samples in which they are observed, they are seen in all regions of the samples imaged, but with varying orientations, periods, and amplitudes. Individual striae appear to be continuous over distances of at least 81 $\mu$m (see e.g. Fig. \ref{fig:suscscanz}d). 
The striae persist, with little variation in period, as the temperature is increased, up to close to the critical temperature. An example is shown in Fig. \ref{fig:mag_susc_vs_T}. In this figure simultaneously taken magnetometry images are in the left column, and susceptibility images are in the right column. The data were taken by cooling in a field of 46 $\mu$T, then imaged at successively higher temperatures as labeled. The dashed lines in the susceptibility images show the positions of cross-sections displayed in Fig. \ref{fig:cross_vs_T}a. 

Cross-sections through the data of Fig. \ref{fig:mag_susc_vs_T}, displayed in Fig. \ref{fig:cross_vs_T}a, show that although there are large, temperature dependent background features in the susceptibility, and the amplitude of the striae fall off as T$_c$ is approached, there is little dependence of the striae periods with temperature. In an effort to better quantify the temperature dependence, we performed two-dimensional Fourier transforms of our susceptibility images. As demonstrated in Fig. \ref{fig:cross_vs_T}b, these transformed images display sharp peaks corresponding to the striae. Fig. \ref{fig:cross_vs_T}c plots the amplitudes, and Fig. \ref{fig:cross_vs_T}d plots the periods, of these peaks as a function of temperature. A comparison with the temperature dependence of the Pearl lengths displayed in Fig. \ref{fig:RLambdafits} shows that the striae amplitudes and periods are less sensitive to temperature than the Pearl lengths. 

\begin{figure}
\centering
\includegraphics[width=\columnwidth]{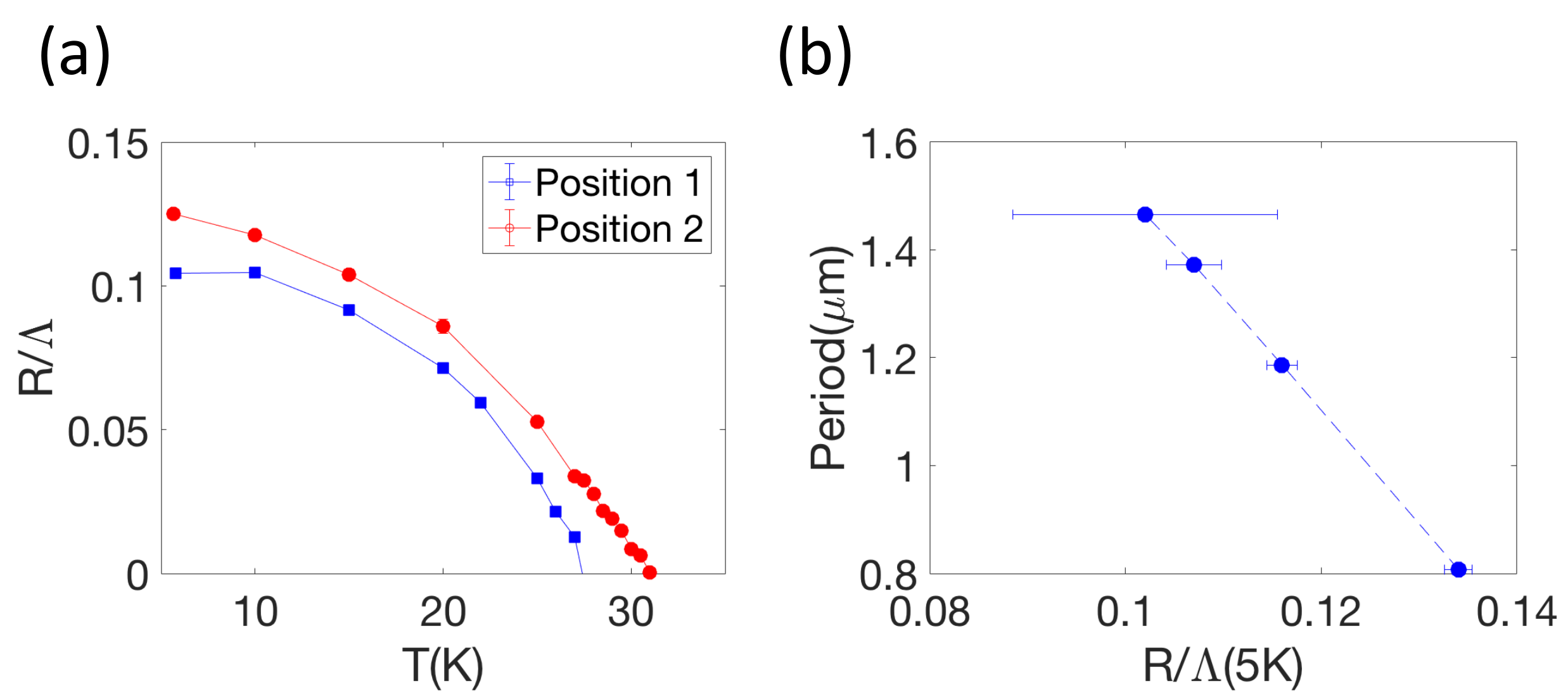}
\caption{{\bf Stripe period varies with low temperature Pearl length.} (a) Fit values from susceptibility touchdown curves for AC174 at two positions separated by 1~mm. (b) Plot of striae period vs. fit values for $R/\Lambda(5K)$ at several positions in sample AC174. The dots are data and the dashed lines connect the dots.}
\label{fig:stripes_vs_position}
\end{figure}

Although the striae periods depend weakly on temperature, there is a strong relation between striae period and low temperature Pearl length. In sample AC 174 (nominal $x_{\rm nom}$=0.125) the critical temperature, striae period, and low temperature Pearl length all vary from position to position on the sample. An example is displayed in 
Fig.~\ref{fig:stripes_vs_position}. Fig.~\ref{fig:stripes_vs_position}a plots the Pearl lengths at two positions separated by 1 mm. Figure~\ref{fig:stripes_vs_position}b plots the stripe period vs R/$\Lambda(T=5K)$ fit values for four positions, covering a range of 0.7 mm on the sample. This figure shows that there is a nearly linear relation between the stripe period and the low temperature inverse Pearl length. This is at first surprising, since the temperature dependent measurements of e.g. Fig. \ref{fig:cross_vs_T} show little dependence of the stripe period on temperature, while our touchdown data e.g. Fig. \ref{fig:RLambdafits} shows that the Pearl length varies strongly with temperature. This may imply that whatever controls the period, e.g. strain, also modulates the carrier density.

A final clue to the mechanism for striae formation in these samples is their orientation. We found that the stripes do not necessarily align with the crystalline axes, the orientations of the striae vary smoothly from position to position on the sample, and we have never observed boundaries between striae with different orientations.
\begin{figure}[h!]
\centering
\includegraphics[width=3in]{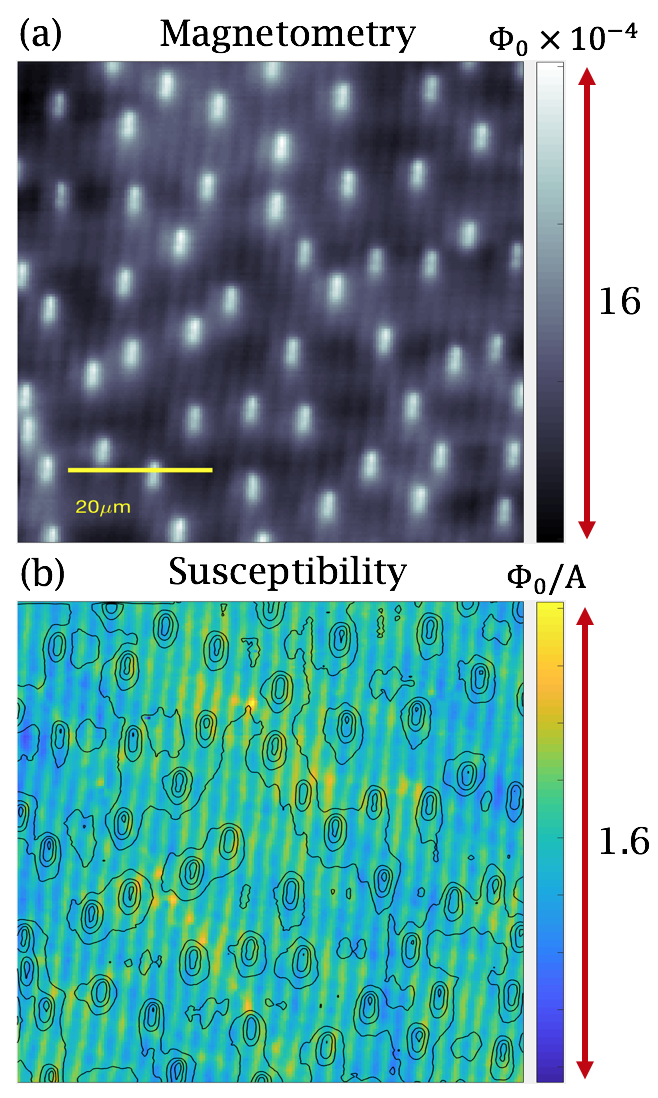}
\caption{{\bf Superconducting vortices trap at positions of largest diamagnetic susceptibility.} Simultaneously acquired magnetometry (upper) and susceptibility (lower) images of sample AC 174, cooled in a field of 23 $\mu$T and imaged in field at 5 K. Superposed on the susceptibility image are five equally spaced contours of magnetic flux (in black).}
\label{fig:contours}
\end{figure}

\begin{figure}[h!]
\centering
\includegraphics[height=6in]{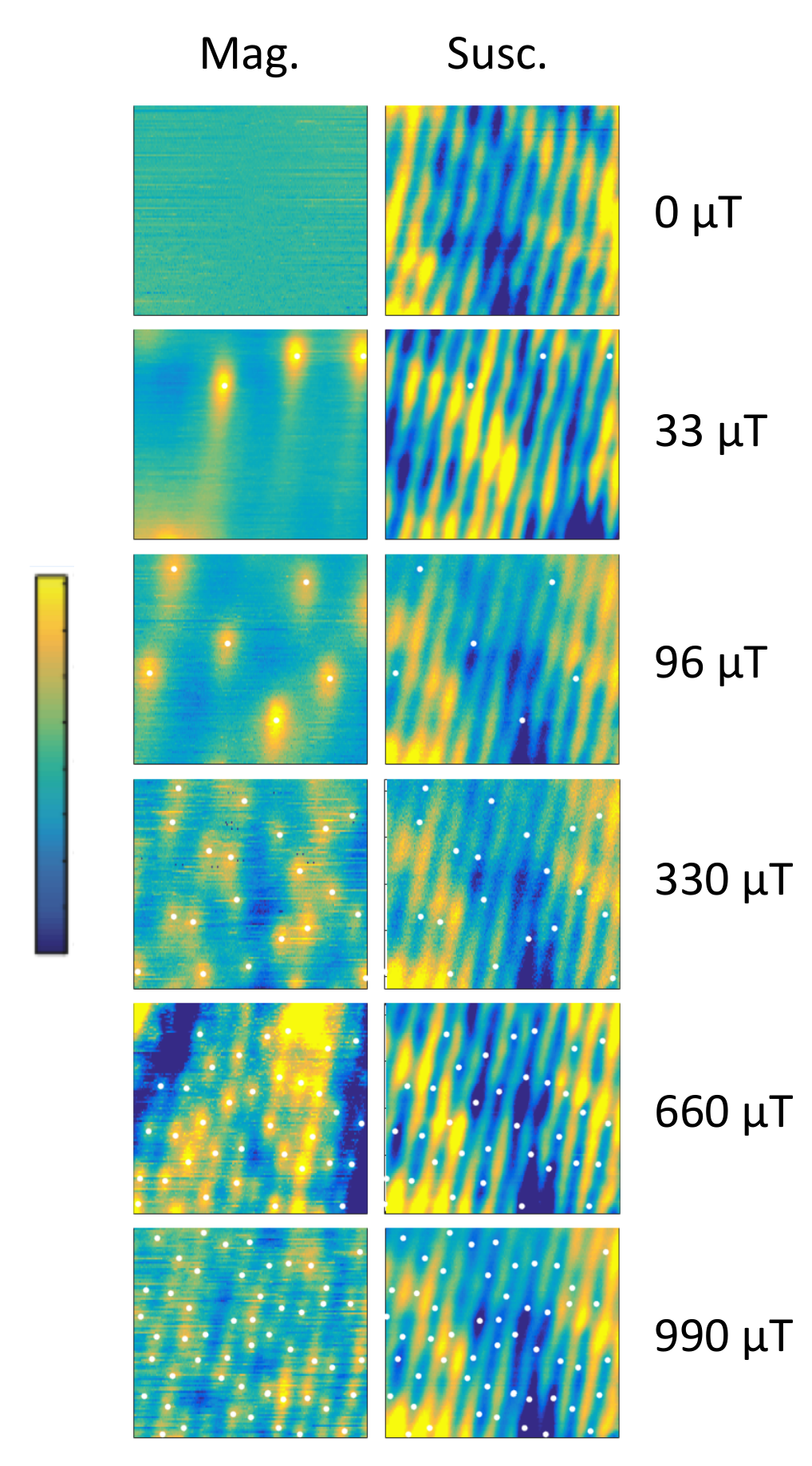}
\caption{{\bf Vortices trap in regions of largest diamagnetic susceptibility, independent of cooling field}. Magnetometry (left column) and susceptibility (right column) images of a 10~$\mu$m wide by 9~$\mu$m high area of sample AC 174, imaged at T=5~K with a field coil modulation of 10~mA at 2.204~kHz. The samples were cooled and imaged in the fields as labeled. The false color scales are 3~m$\Phi_0$ for the magnetometry images, and 0.1~$\Phi_0$/A for the susceptibility images. The white dots represent the centers of the vortices, offset by -0.25 $\mu$m in the susceptibility images to compensate for the displacement between the maximum magnetometry and susceptibility sensitivities for our SQUID sensor. }
\label{fig:mag_susc_vs_B}
\end{figure}

\subsection{Vortex trapping}
Another surprising result is that simultaneous magnetometry and susceptibility images show that superconducting vortices trap on local maxima of the diamagnetic susceptibility. An example of this is shown in Fig. \ref{fig:contours}. The white dots in the magnetometry image in this figure are superconducting vortices trapped when the sample AC~174 (nominal $x_{\rm nom}$=0.125) is cooled in a field of approximately 23~$\mu$T. The vortex images are elongated by the point spread function of the susceptometer used.\cite{kirtley2012scanning} Superimposed on the false-color susceptibility images are contour plots of the magnetometry data, which show that the peaks in the vortex magnetic fields line up with the most diamagnetic regions in the susceptibility images.

This striking effect occurs independently of the cooling field used. An example is shown in Fig. \ref{fig:mag_susc_vs_B}. In this figure the left column shows magnetometry data, and the right column shows susceptibility data. Each row in this figure corresponds to simultaneously acquired data, taken after the sample is cooled in fields as labeled. The centers of the vortices are labeled with white dots in both the magnetometry images and the susceptibility images, showing that the vortices are trapped where the diamagnetic susceptibility is largest, for fields from zero up to 1000 $\mu$T.

\section{Discussion}
We report here stripe-like modulations (striae) in the susceptibility of four samples (out of five measured) of thin film La$_{2-x}$Ba$_{x}$CuO$_{4}$ with nominal dopings of $0.090 < x_{\rm nom} < 0.135$. The striae have periods between one and four microns, approximately a thousand times larger than the periods of stripe phases such as the spin and charge density waves that have been measured in bulk samples using neutron and x-ray diffraction techniques.\cite{Fujita2004,HueckerPRB2011,Tranquada2014} The striae periods are, however, comparable to the Pearl lengths in our films.
The observed modulations in susceptibility are up to eight percent of the full scale. Strikingly, we found that magnetic vortices tend to pin on the diamagnetic maxima of the striae, rather than the minima. The orientation of the striae relative to the crystal axes vary from sample to sample, and from position to position in the same sample.

What causes the striae? We do not believe that they are caused by modulations in film thicknesses, which would cause oscillations in our observed Pearl lengths. Room temperature AFM measurements (Fig. \ref{fig:AFM}b show that the films are flat on the nm scale aside from randomly scattered 20 nm high bumps. However, our measurements are at low temperature. We know that there is a phase transition between the two temperatures, so buckling is possible. Even so, an oscillation in film thickness would not be expected to cause the vortices to trap on maxima of the diamagnetic susceptibility. 

Stripe-like patterns have been observed in many systems, such as materials that undergo martensitic phase transitions and thin film type I superconductors. 
Martensitic phase transitions are diffusionless structural transformations of crystalline materials into highly strained lattice structures. Structural changes result from homogeneous lattice deformations, which are commonly driven by quenching or applied stress.\cite{martensite} To minimize the net strain over large length scales, martensites form elaborate morphologies, such as tweed microstructures, which have domains of unilateral strain along distinct crystal axes.\cite{sethnaStatMech} Tetragonal tweed phases have been observed in cuprate superconductors such as YBCO single crystals.\cite{Liang1995} Since experiments have shown that epitaxial strain can enhance local superconductivity in cuprates,\cite{Locquet1998} periodic, stripe-like domains of uniaxial  strain in a martensite may induce the modulations in the Pearl length that characterize the striae. Moreover, similar stripe-like domains of alternating superconducting and normal regions have been observed in thin film type I superconductors.\cite{Seul1995} In this case, applied magnetic fields drive the the structural separation of superconducting and normal phases. 

Based on numerous previous studies on the pinning of superconducting vortices, particularly in cuprate superconductors, our observation that the vortices pin in the regions where the diamagnetic susceptibility is higher is surprising. In general, the total energy of the vortex is the energy of the superfluid part plus the energy of the core.\cite{blatter1994vortices} The energy of the superfluid part of the vortex must be higher in a region with higher diamagnetism,\cite{pearl1964current, carneiro2000vortex} and generally one would expect this to be true for the core energy as well. Since we expect that vortices would be attracted to regions of reduced superfluid density, the vortex core energy in our samples must be different than expected. Given  the rich competition of order parameters in the superconducting phase of La$_{2-x}$Ba$_{x}$CuO$_4$, it is plausible that energy contributions from sources such as strain and pair density waves to the free energy of the system render it energetically favorable for the vortices to pin on diamagnetic maxima. One model for pair density waves in superconducting vortex halos proposes that the order parameter of a pair density wave peaks at vortex cores.\cite{Wang2018} If, for example, the condensation energy that is gained from a pair density wave is close to the condensation energy that is lost by superconductivity, the vortex core energies could be minimized at peaks of superfluid density.

\section{Acknowledgements}
This work was primarily supported by the Department of Energy, Office of Basic Energy Sciences, Division of Materials Sciences and Engineering, under Contract No. DE-AC02-76SF00515. The AFM measurements were done by Andrey V. Malkovskiy and the XPS measurements by Chuck Hitzman at the Stanford Nano Shared Facilities (SNSF), supported by the National Science Foundation under award ECCS-1542152. The scanning SQUID microscope and sensors used in this experiment were developed with support by an NSF IMR-MIP Grant No. DMR-0957616. C.A. was supported by an Air Force Office of Scientific Research grant No. FA9550-09-1-0583.

\bibliography{references}

\begin{thebibliography}{29}%
\makeatletter
\providecommand \@ifxundefined [1]{%
 \@ifx{#1\undefined}
}%
\providecommand \@ifnum [1]{%
 \ifnum #1\expandafter \@firstoftwo
 \else \expandafter \@secondoftwo
 \fi
}%
\providecommand \@ifx [1]{%
 \ifx #1\expandafter \@firstoftwo
 \else \expandafter \@secondoftwo
 \fi
}%
\providecommand \natexlab [1]{#1}%
\providecommand \enquote  [1]{``#1''}%
\providecommand \bibnamefont  [1]{#1}%
\providecommand \bibfnamefont [1]{#1}%
\providecommand \citenamefont [1]{#1}%
\providecommand \href@noop [0]{\@secondoftwo}%
\providecommand \href [0]{\begingroup \@sanitize@url \@href}%
\providecommand \@href[1]{\@@startlink{#1}\@@href}%
\providecommand \@@href[1]{\endgroup#1\@@endlink}%
\providecommand \@sanitize@url [0]{\catcode `\\12\catcode `\$12\catcode
  `\&12\catcode `\#12\catcode `\^12\catcode `\_12\catcode `\%12\relax}%
\providecommand \@@startlink[1]{}%
\providecommand \@@endlink[0]{}%
\providecommand \url  [0]{\begingroup\@sanitize@url \@url }%
\providecommand \@url [1]{\endgroup\@href {#1}{\urlprefix }}%
\providecommand \urlprefix  [0]{URL }%
\providecommand \Eprint [0]{\href }%
\providecommand \doibase [0]{http://dx.doi.org/}%
\providecommand \selectlanguage [0]{\@gobble}%
\providecommand \bibinfo  [0]{\@secondoftwo}%
\providecommand \bibfield  [0]{\@secondoftwo}%
\providecommand \translation [1]{[#1]}%
\providecommand \BibitemOpen [0]{}%
\providecommand \bibitemStop [0]{}%
\providecommand \bibitemNoStop [0]{.\EOS\space}%
\providecommand \EOS [0]{\spacefactor3000\relax}%
\providecommand \BibitemShut  [1]{\csname bibitem#1\endcsname}%
\let\auto@bib@innerbib\@empty
\bibitem [{\citenamefont {Bednorz}\ and\ \citenamefont
  {Muller}(1986)}]{BednorzPRB86}%
  \BibitemOpen
  \bibfield  {author} {\bibinfo {author} {\bibfnamefont {J.G.}\ \bibnamefont
  {Bednorz}}\ and\ \bibinfo {author} {\bibfnamefont {K.A.}\ \bibnamefont
  {Muller}},\ }\bibfield  {title} {\enquote {\bibinfo {title} {Possible {H}igh
  {T}$_c$ {S}uperconductivity in the {Ba–La–Cu–O} system},}\ }\href@noop
  {} {\bibfield  {journal} {\bibinfo  {journal} {Z. Phys. B Condensed Matter}\
  }\textbf {\bibinfo {volume} {64}},\ \bibinfo {pages} {189--193} (\bibinfo
  {year} {1986})}\BibitemShut {NoStop}%
\bibitem [{\citenamefont {Tsuei}\ and\ \citenamefont
  {Kirtley}(2000)}]{Tsuei2000}%
  \BibitemOpen
  \bibfield  {author} {\bibinfo {author} {\bibfnamefont {C.~C.}\ \bibnamefont
  {Tsuei}}\ and\ \bibinfo {author} {\bibfnamefont {J.~R.}\ \bibnamefont
  {Kirtley}},\ }\bibfield  {title} {\enquote {\bibinfo {title} {Pairing
  symmetry in cuprate superconductors},}\ }\href@noop {} {\bibfield  {journal}
  {\bibinfo  {journal} {Rev. {M}od. {P}hys}\ }\textbf {\bibinfo {volume}
  {72}},\ \bibinfo {pages} {3377–3395} (\bibinfo {year} {2000})}\BibitemShut
  {NoStop}%
\bibitem [{\citenamefont {Norman}\ and\ \citenamefont
  {Pepin}(2003)}]{norman2003electronic}%
  \BibitemOpen
  \bibfield  {author} {\bibinfo {author} {\bibfnamefont {MR}~\bibnamefont
  {Norman}}\ and\ \bibinfo {author} {\bibfnamefont {C}~\bibnamefont {Pepin}},\
  }\bibfield  {title} {\enquote {\bibinfo {title} {The electronic nature of
  high temperature cuprate superconductors},}\ }\href@noop {} {\bibfield
  {journal} {\bibinfo  {journal} {Reports on {P}rogress in {P}hysics}\ }\textbf
  {\bibinfo {volume} {66}},\ \bibinfo {pages} {1547} (\bibinfo {year}
  {2003})}\BibitemShut {NoStop}%
\bibitem [{\citenamefont {Lee}(2007)}]{lee2007high}%
  \BibitemOpen
  \bibfield  {author} {\bibinfo {author} {\bibfnamefont {Patrick~A}\
  \bibnamefont {Lee}},\ }\bibfield  {title} {\enquote {\bibinfo {title} {From
  high temperature superconductivity to quantum spin liquid: {P}rogress in
  strong correlation physics},}\ }\href@noop {} {\bibfield  {journal} {\bibinfo
   {journal} {Reports on {P}rogress in {P}hysics}\ }\textbf {\bibinfo {volume}
  {71}},\ \bibinfo {pages} {012501} (\bibinfo {year} {2007})}\BibitemShut
  {NoStop}%
\bibitem [{\citenamefont {Lee}\ \emph {et~al.}(2017)\citenamefont {Lee},
  \citenamefont {Yoon},\ and\ \citenamefont {Abd-Shukor}}]{LeeJSNM2017}%
  \BibitemOpen
  \bibfield  {author} {\bibinfo {author} {\bibfnamefont {B.S.}\ \bibnamefont
  {Lee}}, \bibinfo {author} {\bibfnamefont {T.L.}\ \bibnamefont {Yoon}}, \ and\
  \bibinfo {author} {\bibfnamefont {R.}~\bibnamefont {Abd-Shukor}},\ }\bibfield
   {title} {\enquote {\bibinfo {title} {Theory of d-wave high temperature
  superconductivity in the cuprates involving non-linear lattice modes},}\
  }\href@noop {} {\bibfield  {journal} {\bibinfo  {journal} {Journal of
  {S}uperconductivity and {N}ovel {M}agnetism}\ }\textbf {\bibinfo {volume}
  {30}},\ \bibinfo {pages} {3377–3395} (\bibinfo {year} {2017})}\BibitemShut
  {NoStop}%
\bibitem [{\citenamefont {Nie}\ \emph {et~al.}(2014)\citenamefont {Nie},
  \citenamefont {Tarjus},\ and\ \citenamefont {Kivelson}}]{nie2014quenched}%
  \BibitemOpen
  \bibfield  {author} {\bibinfo {author} {\bibfnamefont {Laimei}\ \bibnamefont
  {Nie}}, \bibinfo {author} {\bibfnamefont {Gilles}\ \bibnamefont {Tarjus}}, \
  and\ \bibinfo {author} {\bibfnamefont {Steven~Allan}\ \bibnamefont
  {Kivelson}},\ }\bibfield  {title} {\enquote {\bibinfo {title} {Quenched
  disorder and vestigial nematicity in the pseudogap regime of the cuprates},}\
  }\href@noop {} {\bibfield  {journal} {\bibinfo  {journal} {Proceedings of the
  {N}ational {A}cademy of {S}ciences}\ }\textbf {\bibinfo {volume} {111}},\
  \bibinfo {pages} {7980--7985} (\bibinfo {year} {2014})}\BibitemShut {NoStop}%
\bibitem [{\citenamefont {Berg}\ \emph {et~al.}(2009)\citenamefont {Berg},
  \citenamefont {Fradkin},\ and\ \citenamefont {Kivelson}}]{Berg2009charge}%
  \BibitemOpen
  \bibfield  {author} {\bibinfo {author} {\bibfnamefont {Erez}\ \bibnamefont
  {Berg}}, \bibinfo {author} {\bibfnamefont {Eduardo}\ \bibnamefont {Fradkin}},
  \ and\ \bibinfo {author} {\bibfnamefont {Steven~A}\ \bibnamefont
  {Kivelson}},\ }\bibfield  {title} {\enquote {\bibinfo {title} {Charge-4e
  superconductivity from pair-density-wave order in certain high-temperature
  superconductors},}\ }\href@noop {} {\bibfield  {journal} {\bibinfo  {journal}
  {Nat. Phys.}\ }\textbf {\bibinfo {volume} {5}},\ \bibinfo {pages} {830}
  (\bibinfo {year} {2009})}\BibitemShut {NoStop}%
\bibitem [{\citenamefont {Fujita}\ \emph {et~al.}(2002)\citenamefont {Fujita},
  \citenamefont {Goka}, \citenamefont {Yamada},\ and\ \citenamefont
  {Matsuda}}]{fujita2002competition}%
  \BibitemOpen
  \bibfield  {author} {\bibinfo {author} {\bibfnamefont {M}~\bibnamefont
  {Fujita}}, \bibinfo {author} {\bibfnamefont {H}~\bibnamefont {Goka}},
  \bibinfo {author} {\bibfnamefont {K}~\bibnamefont {Yamada}}, \ and\ \bibinfo
  {author} {\bibfnamefont {M}~\bibnamefont {Matsuda}},\ }\bibfield  {title}
  {\enquote {\bibinfo {title} {Competition between {Charge}-and
  {Spin-Density-Wave} {Order} and {Superconductivity} in
  {La$_{1.875}$Ba$_{0.125-x}$Sr$_x$CuO$_4$}},}\ }\href@noop {} {\bibfield
  {journal} {\bibinfo  {journal} {Physical {R}eview {L}etters}\ }\textbf
  {\bibinfo {volume} {88}},\ \bibinfo {pages} {167008} (\bibinfo {year}
  {2002})}\BibitemShut {NoStop}%
\bibitem [{\citenamefont {Kirtley}(2010)}]{kirtley2010fundamental}%
  \BibitemOpen
  \bibfield  {author} {\bibinfo {author} {\bibfnamefont {JR}~\bibnamefont
  {Kirtley}},\ }\bibfield  {title} {\enquote {\bibinfo {title} {Fundamental
  studies of superconductors using scanning magnetic imaging},}\ }\href@noop {}
  {\bibfield  {journal} {\bibinfo  {journal} {Rep. Prog. Phys.}\ }\textbf
  {\bibinfo {volume} {73}},\ \bibinfo {pages} {126501} (\bibinfo {year}
  {2010})}\BibitemShut {NoStop}%
\bibitem [{\citenamefont {Kirtley}\ \emph {et~al.}(2012)\citenamefont
  {Kirtley}, \citenamefont {Kalisky}, \citenamefont {Bert}, \citenamefont
  {Bell}, \citenamefont {Kim}, \citenamefont {Hikita}, \citenamefont {Hwang},
  \citenamefont {Ngai}, \citenamefont {Segal}, \citenamefont {Walker} \emph
  {et~al.}}]{kirtley2012scanning}%
  \BibitemOpen
  \bibfield  {author} {\bibinfo {author} {\bibfnamefont {JR}~\bibnamefont
  {Kirtley}}, \bibinfo {author} {\bibfnamefont {B}~\bibnamefont {Kalisky}},
  \bibinfo {author} {\bibfnamefont {JA}~\bibnamefont {Bert}}, \bibinfo {author}
  {\bibfnamefont {C}~\bibnamefont {Bell}}, \bibinfo {author} {\bibfnamefont
  {M}~\bibnamefont {Kim}}, \bibinfo {author} {\bibfnamefont {Y}~\bibnamefont
  {Hikita}}, \bibinfo {author} {\bibfnamefont {HY}~\bibnamefont {Hwang}},
  \bibinfo {author} {\bibfnamefont {JH}~\bibnamefont {Ngai}}, \bibinfo {author}
  {\bibfnamefont {Y}~\bibnamefont {Segal}}, \bibinfo {author} {\bibfnamefont
  {FJ}~\bibnamefont {Walker}},  \emph {et~al.},\ }\bibfield  {title} {\enquote
  {\bibinfo {title} {Scanning {SQUID} susceptometry of a paramagnetic
  superconductor},}\ }\href@noop {} {\bibfield  {journal} {\bibinfo  {journal}
  {Physical Review B}\ }\textbf {\bibinfo {volume} {85}},\ \bibinfo {pages}
  {224518} (\bibinfo {year} {2012})}\BibitemShut {NoStop}%
\bibitem [{\citenamefont {Kirtley}\ \emph {et~al.}(1995)\citenamefont
  {Kirtley}, \citenamefont {Ketchen}, \citenamefont {Stawiasz}, \citenamefont
  {Sun}, \citenamefont {Gallagher}, \citenamefont {Blanton},\ and\
  \citenamefont {Wind}}]{kirtley1995high}%
  \BibitemOpen
  \bibfield  {author} {\bibinfo {author} {\bibfnamefont {JR}~\bibnamefont
  {Kirtley}}, \bibinfo {author} {\bibfnamefont {MB}~\bibnamefont {Ketchen}},
  \bibinfo {author} {\bibfnamefont {KG}~\bibnamefont {Stawiasz}}, \bibinfo
  {author} {\bibfnamefont {JZ}~\bibnamefont {Sun}}, \bibinfo {author}
  {\bibfnamefont {WJ}~\bibnamefont {Gallagher}}, \bibinfo {author}
  {\bibfnamefont {SH}~\bibnamefont {Blanton}}, \ and\ \bibinfo {author}
  {\bibfnamefont {SJ}~\bibnamefont {Wind}},\ }\bibfield  {title} {\enquote
  {\bibinfo {title} {High-resolution scanning {SQUID} microscope},}\
  }\href@noop {} {\bibfield  {journal} {\bibinfo  {journal} {Applied {P}hysics
  {L}etters}\ }\textbf {\bibinfo {volume} {66}},\ \bibinfo {pages} {1138--1140}
  (\bibinfo {year} {1995})}\BibitemShut {NoStop}%
\bibitem [{\citenamefont {Gardner}\ \emph {et~al.}(2001)\citenamefont
  {Gardner}, \citenamefont {Wynn}, \citenamefont {Bj{\"o}rnsson}, \citenamefont
  {Straver}, \citenamefont {Moler}, \citenamefont {Kirtley},\ and\
  \citenamefont {Ketchen}}]{gardner2001scanning}%
  \BibitemOpen
  \bibfield  {author} {\bibinfo {author} {\bibfnamefont {Brian~W}\ \bibnamefont
  {Gardner}}, \bibinfo {author} {\bibfnamefont {Janice~C}\ \bibnamefont
  {Wynn}}, \bibinfo {author} {\bibfnamefont {Per~G}\ \bibnamefont
  {Bj{\"o}rnsson}}, \bibinfo {author} {\bibfnamefont {Eric~WJ}\ \bibnamefont
  {Straver}}, \bibinfo {author} {\bibfnamefont {Kathryn~A}\ \bibnamefont
  {Moler}}, \bibinfo {author} {\bibfnamefont {John~R}\ \bibnamefont {Kirtley}},
  \ and\ \bibinfo {author} {\bibfnamefont {Mark~B}\ \bibnamefont {Ketchen}},\
  }\bibfield  {title} {\enquote {\bibinfo {title} {Scanning superconducting
  quantum interference device susceptometry},}\ }\href@noop {} {\bibfield
  {journal} {\bibinfo  {journal} {Review of {S}cientific {I}nstruments}\
  }\textbf {\bibinfo {volume} {72}},\ \bibinfo {pages} {2361--2364} (\bibinfo
  {year} {2001})}\BibitemShut {NoStop}%
\bibitem [{\citenamefont {Kirtley}\ \emph {et~al.}(2016)\citenamefont
  {Kirtley}, \citenamefont {Paulius}, \citenamefont {Rosenberg}, \citenamefont
  {Palmstrom}, \citenamefont {Holland}, \citenamefont {Spanton}, \citenamefont
  {Schiessl}, \citenamefont {Jermain}, \citenamefont {Gibbons}, \citenamefont
  {Fung} \emph {et~al.}}]{kirtley2016scanning}%
  \BibitemOpen
  \bibfield  {author} {\bibinfo {author} {\bibfnamefont {John~R}\ \bibnamefont
  {Kirtley}}, \bibinfo {author} {\bibfnamefont {Lisa}\ \bibnamefont {Paulius}},
  \bibinfo {author} {\bibfnamefont {Aaron~J}\ \bibnamefont {Rosenberg}},
  \bibinfo {author} {\bibfnamefont {Johanna~C}\ \bibnamefont {Palmstrom}},
  \bibinfo {author} {\bibfnamefont {Connor~M}\ \bibnamefont {Holland}},
  \bibinfo {author} {\bibfnamefont {Eric~M}\ \bibnamefont {Spanton}}, \bibinfo
  {author} {\bibfnamefont {Daniel}\ \bibnamefont {Schiessl}}, \bibinfo {author}
  {\bibfnamefont {Colin~L}\ \bibnamefont {Jermain}}, \bibinfo {author}
  {\bibfnamefont {Jonathan}\ \bibnamefont {Gibbons}}, \bibinfo {author}
  {\bibfnamefont {Y-K-K}\ \bibnamefont {Fung}},  \emph {et~al.},\ }\bibfield
  {title} {\enquote {\bibinfo {title} {Scanning {SQUID} susceptometers with
  sub-micron spatial resolution},}\ }\href@noop {} {\bibfield  {journal}
  {\bibinfo  {journal} {Review of {S}cientific {I}nstruments}\ }\textbf
  {\bibinfo {volume} {87}},\ \bibinfo {pages} {093702} (\bibinfo {year}
  {2016})}\BibitemShut {NoStop}%
\bibitem [{\citenamefont {Haeni}\ \emph {et~al.}(2000)\citenamefont {Haeni},
  \citenamefont {Theis},\ and\ \citenamefont {Schlom}}]{haeni2000RHEED}%
  \BibitemOpen
  \bibfield  {author} {\bibinfo {author} {\bibfnamefont {J.H.}\ \bibnamefont
  {Haeni}}, \bibinfo {author} {\bibfnamefont {C.D.}\ \bibnamefont {Theis}}, \
  and\ \bibinfo {author} {\bibfnamefont {D.G.}\ \bibnamefont {Schlom}},\
  }\bibfield  {title} {\enquote {\bibinfo {title} {{RHEED} {I}ntensity
  {O}scillations for the {S}toichiometric {G}rowth of {S}r{T}i{O}$_3$ {T}hin
  {F}ilms by {R}eactive {M}olecular {B}eam {E}pitaxy},}\ }\href@noop {}
  {\bibfield  {journal} {\bibinfo  {journal} {J. Electroceram.}\ }\textbf
  {\bibinfo {volume} {4}},\ \bibinfo {pages} {385--391} (\bibinfo {year}
  {2000})}\BibitemShut {NoStop}%
\bibitem [{att()}]{attocube}%
  \BibitemOpen
  \href@noop {} {\enquote {\bibinfo {title} {{Attocube scanning {SQUID}
  platform}},}\ }\bibinfo {howpublished}
  {http://www.attocube.com/attomicroscopy/microscopy-solutions/squid/},\
  \bibinfo {note} {[Accessed: 28-January-2018]}\BibitemShut {NoStop}%
\bibitem [{\citenamefont {Kirtley}\ \emph {et~al.}(1999)\citenamefont
  {Kirtley}, \citenamefont {Tsuei}, \citenamefont {Moler}, \citenamefont
  {Kogan}, \citenamefont {Clem},\ and\ \citenamefont
  {Turberfield}}]{kirtley1999variable}%
  \BibitemOpen
  \bibfield  {author} {\bibinfo {author} {\bibfnamefont {JR}~\bibnamefont
  {Kirtley}}, \bibinfo {author} {\bibfnamefont {CC}~\bibnamefont {Tsuei}},
  \bibinfo {author} {\bibfnamefont {KA}~\bibnamefont {Moler}}, \bibinfo
  {author} {\bibfnamefont {VG}~\bibnamefont {Kogan}}, \bibinfo {author}
  {\bibfnamefont {JR}~\bibnamefont {Clem}}, \ and\ \bibinfo {author}
  {\bibfnamefont {AJ}~\bibnamefont {Turberfield}},\ }\bibfield  {title}
  {\enquote {\bibinfo {title} {Variable sample temperature scanning
  superconducting quantum interference device microscope},}\ }\href@noop {}
  {\bibfield  {journal} {\bibinfo  {journal} {Applied {P}hysics {L}etters}\
  }\textbf {\bibinfo {volume} {74}},\ \bibinfo {pages} {4011--4013} (\bibinfo
  {year} {1999})}\BibitemShut {NoStop}%
\bibitem [{\citenamefont {Sato}\ \emph {et~al.}(2000)\citenamefont {Sato},
  \citenamefont {Tsukada}, \citenamefont {Naito},\ and\ \citenamefont
  {Matsuda}}]{sato2000absence}%
  \BibitemOpen
  \bibfield  {author} {\bibinfo {author} {\bibfnamefont {H.}~\bibnamefont
  {Sato}}, \bibinfo {author} {\bibfnamefont {A.}~\bibnamefont {Tsukada}},
  \bibinfo {author} {\bibfnamefont {M.}~\bibnamefont {Naito}}, \ and\ \bibinfo
  {author} {\bibfnamefont {A.}~\bibnamefont {Matsuda}},\ }\bibfield  {title}
  {\enquote {\bibinfo {title} {Absence of 1/8 anomaly in strained thin films of
  {La$_{2-x}$Ba$_x$CuO$_{4+\delta}$}},}\ }\href@noop {} {\bibfield  {journal}
  {\bibinfo  {journal} {Physical {R}eview {B}}\ }\textbf {\bibinfo {volume}
  {62}},\ \bibinfo {pages} {R799} (\bibinfo {year} {2000})}\BibitemShut
  {NoStop}%
\bibitem [{\citenamefont {Fujita}\ \emph {et~al.}(2004)\citenamefont {Fujita},
  \citenamefont {Goka}, \citenamefont {Yamada}, \citenamefont {Tranquada},\
  and\ \citenamefont {Regnault}}]{Fujita2004}%
  \BibitemOpen
  \bibfield  {author} {\bibinfo {author} {\bibfnamefont {M.}~\bibnamefont
  {Fujita}}, \bibinfo {author} {\bibfnamefont {H.}~\bibnamefont {Goka}},
  \bibinfo {author} {\bibfnamefont {K.}~\bibnamefont {Yamada}}, \bibinfo
  {author} {\bibfnamefont {J.~M.}\ \bibnamefont {Tranquada}}, \ and\ \bibinfo
  {author} {\bibfnamefont {L.~P.}\ \bibnamefont {Regnault}},\ }\bibfield
  {title} {\enquote {\bibinfo {title} {Stripe order, depinning, and
  fluctuations in {L}a$_{1.875}${B}a$_{0.125}${C}uo$_4$ and
  {L}a$_{1.875}${B}a$_{0.075}${S}r$_{0.050}${C}u{O}$_4$},}\ }\href@noop {}
  {\bibfield  {journal} {\bibinfo  {journal} {Physical {R}eview {B}}\ }\textbf
  {\bibinfo {volume} {70}},\ \bibinfo {pages} {104517} (\bibinfo {year}
  {2004})}\BibitemShut {NoStop}%
\bibitem [{\citenamefont {Huecker}\ \emph {et~al.}(2011)\citenamefont
  {Huecker}, \citenamefont {Zimmermann}, \citenamefont {Gu}, \citenamefont
  {Xu}, \citenamefont {Wen}, \citenamefont {Xu}, \citenamefont {Kang},
  \citenamefont {Zheludev},\ and\ \citenamefont {M.}}]{HueckerPRB2011}%
  \BibitemOpen
  \bibfield  {author} {\bibinfo {author} {\bibfnamefont {M.}~\bibnamefont
  {Huecker}}, \bibinfo {author} {\bibfnamefont {M.~V.}\ \bibnamefont
  {Zimmermann}}, \bibinfo {author} {\bibfnamefont {G.D.}\ \bibnamefont {Gu}},
  \bibinfo {author} {\bibfnamefont {Z.~J.}\ \bibnamefont {Xu}}, \bibinfo
  {author} {\bibfnamefont {J.~S.}\ \bibnamefont {Wen}}, \bibinfo {author}
  {\bibfnamefont {Guanyong}\ \bibnamefont {Xu}}, \bibinfo {author}
  {\bibfnamefont {H.~J.}\ \bibnamefont {Kang}}, \bibinfo {author}
  {\bibfnamefont {A.}~\bibnamefont {Zheludev}}, \ and\ \bibinfo {author}
  {\bibfnamefont {Tranquada~J.}\ \bibnamefont {M.}},\ }\bibfield  {title}
  {\enquote {\bibinfo {title} {Stripe order in superconducting
  {L}a$_{2-x}${B}a$_x${C}uo$_4$ for 0.095 $<=$ x $<=$ 0.155},}\ }\href@noop {}
  {\bibfield  {journal} {\bibinfo  {journal} {Phys. Rev. B}\ }\textbf {\bibinfo
  {volume} {83}},\ \bibinfo {pages} {104506} (\bibinfo {year}
  {2011})}\BibitemShut {NoStop}%
\bibitem [{\citenamefont {Tranquada}(2014)}]{Tranquada2014}%
  \BibitemOpen
  \bibfield  {author} {\bibinfo {author} {\bibfnamefont {J.~M.}\ \bibnamefont
  {Tranquada}},\ }\bibfield  {title} {\enquote {\bibinfo {title} {Exploring
  intertwined orders in cuprate superconductors},}\ }\href@noop {} {\bibfield
  {journal} {\bibinfo  {journal} {Physica {B}}\ }\textbf {\bibinfo {volume}
  {460}},\ \bibinfo {pages} {136--140} (\bibinfo {year} {2014})}\BibitemShut
  {NoStop}%
\bibitem [{mar()}]{martensite}%
  \BibitemOpen
  \href@noop {} {\enquote {\bibinfo {title} {{IUCr} martensitic phase
  transformations: the memory of shape},}\ }\bibinfo {howpublished}
  {\url{https://www.iucr.org/news/newsletter/volume-7/number-2/martensitic-transformations}},\
  \bibinfo {note} {accessed: 2018-03-15}\BibitemShut {NoStop}%
\bibitem [{\citenamefont {Sethna}(2017)}]{sethnaStatMech}%
  \BibitemOpen
  \bibfield  {author} {\bibinfo {author} {\bibfnamefont {James~P.}\
  \bibnamefont {Sethna}},\ }\href@noop {} {\emph {\bibinfo {title} {Entropy,
  Order Parameters, and Complexity}}}\ (\bibinfo  {publisher} {Clarendon
  Press},\ \bibinfo {address} {Oxford},\ \bibinfo {year} {2017})\ p.\ \bibinfo
  {pages} {250}\BibitemShut {NoStop}%
\bibitem [{\citenamefont {Liang}\ \emph {et~al.}(1995)\citenamefont {Liang},
  \citenamefont {Lin}, \citenamefont {Chrosch}, \citenamefont {Yan},\ and\
  \citenamefont {Salje}}]{Liang1995}%
  \BibitemOpen
  \bibfield  {author} {\bibinfo {author} {\bibfnamefont {W.Y.}\ \bibnamefont
  {Liang}}, \bibinfo {author} {\bibfnamefont {C.T.}\ \bibnamefont {Lin}},
  \bibinfo {author} {\bibfnamefont {J.}~\bibnamefont {Chrosch}}, \bibinfo
  {author} {\bibfnamefont {Y.}~\bibnamefont {Yan}}, \ and\ \bibinfo {author}
  {\bibfnamefont {E.K.H.}\ \bibnamefont {Salje}},\ }\href@noop {} {\bibfield
  {journal} {\bibinfo  {journal} {Advances in {S}uperconductivity {VII}}\
  }\textbf {\bibinfo {volume} {1}},\ \bibinfo {pages} {137--140} (\bibinfo
  {year} {1995})}\BibitemShut {NoStop}%
\bibitem [{\citenamefont {Locquet}\ \emph {et~al.}(1998)\citenamefont
  {Locquet}, \citenamefont {Perret}, \citenamefont {Fompeyrine}, \citenamefont
  {Mächler}, \citenamefont {Seo},\ and\ \citenamefont
  {Tendeloo}}]{Locquet1998}%
  \BibitemOpen
  \bibfield  {author} {\bibinfo {author} {\bibfnamefont {J.-P.}\ \bibnamefont
  {Locquet}}, \bibinfo {author} {\bibfnamefont {J.}~\bibnamefont {Perret}},
  \bibinfo {author} {\bibfnamefont {J.}~\bibnamefont {Fompeyrine}}, \bibinfo
  {author} {\bibfnamefont {E.}~\bibnamefont {Mächler}}, \bibinfo {author}
  {\bibfnamefont {J.~W.}\ \bibnamefont {Seo}}, \ and\ \bibinfo {author}
  {\bibfnamefont {G.~Van}\ \bibnamefont {Tendeloo}},\ }\bibfield  {title}
  {\enquote {\bibinfo {title} {Doubling the critical temperature of
  {L}a$_{1.9}${S}r$_{0.1}${C}u{O}$_4$ using epitaxial strain},}\ }\href@noop {}
  {\bibfield  {journal} {\bibinfo  {journal} {Nature}\ }\textbf {\bibinfo
  {volume} {394}},\ \bibinfo {pages} {453–456} (\bibinfo {year}
  {1998})}\BibitemShut {NoStop}%
\bibitem [{\citenamefont {Seul}\ and\ \citenamefont
  {Andelman}(1998)}]{Seul1995}%
  \BibitemOpen
  \bibfield  {author} {\bibinfo {author} {\bibfnamefont {Michael}\ \bibnamefont
  {Seul}}\ and\ \bibinfo {author} {\bibfnamefont {David}\ \bibnamefont
  {Andelman}},\ }\bibfield  {title} {\enquote {\bibinfo {title} {Domain shapes
  and patterns: The phenomenology of modulated phases},}\ }\href@noop {}
  {\bibfield  {journal} {\bibinfo  {journal} {Science}\ }\textbf {\bibinfo
  {volume} {267}},\ \bibinfo {pages} {476--483} (\bibinfo {year}
  {1998})}\BibitemShut {NoStop}%
\bibitem [{\citenamefont {Blatter}\ \emph {et~al.}(1994)\citenamefont
  {Blatter}, \citenamefont {Feigel'man}, \citenamefont {Geshkenbein},
  \citenamefont {Larkin},\ and\ \citenamefont {Vinokur}}]{blatter1994vortices}%
  \BibitemOpen
  \bibfield  {author} {\bibinfo {author} {\bibfnamefont {Gianni}\ \bibnamefont
  {Blatter}}, \bibinfo {author} {\bibfnamefont {Mikhail~V}\ \bibnamefont
  {Feigel'man}}, \bibinfo {author} {\bibfnamefont {Vadim~B}\ \bibnamefont
  {Geshkenbein}}, \bibinfo {author} {\bibfnamefont {Anatoly~I}\ \bibnamefont
  {Larkin}}, \ and\ \bibinfo {author} {\bibfnamefont {Valerii~M}\ \bibnamefont
  {Vinokur}},\ }\bibfield  {title} {\enquote {\bibinfo {title} {Vortices in
  high-temperature superconductors},}\ }\href@noop {} {\bibfield  {journal}
  {\bibinfo  {journal} {Reviews of Modern Physics}\ }\textbf {\bibinfo {volume}
  {66}},\ \bibinfo {pages} {1125} (\bibinfo {year} {1994})}\BibitemShut
  {NoStop}%
\bibitem [{\citenamefont {Pearl}(1964)}]{pearl1964current}%
  \BibitemOpen
  \bibfield  {author} {\bibinfo {author} {\bibfnamefont {J}~\bibnamefont
  {Pearl}},\ }\bibfield  {title} {\enquote {\bibinfo {title} {Current
  distribution in superconducting films carrying quantized fluxoids},}\
  }\href@noop {} {\bibfield  {journal} {\bibinfo  {journal} {Applied Physics
  Letters}\ }\textbf {\bibinfo {volume} {5}},\ \bibinfo {pages} {65--66}
  (\bibinfo {year} {1964})}\BibitemShut {NoStop}%
\bibitem [{\citenamefont {Carneiro}\ and\ \citenamefont
  {Brandt}()}]{carneiro2000vortex}%
  \BibitemOpen
  \bibfield  {author} {\bibinfo {author} {\bibfnamefont {Gilson}\ \bibnamefont
  {Carneiro}}\ and\ \bibinfo {author} {\bibfnamefont {Ernst~Helmut}\
  \bibnamefont {Brandt}},\ }\bibfield  {title} {\enquote {\bibinfo {title}
  {Vortex lines in films: {F}ields and interactions},}\ }\href@noop {}
  {\bibinfo  {journal} {Physical {R}eview {B}, volume={61}, number={9},
  pages={6370}, year={2000}, publisher={APS}}\ }\BibitemShut {NoStop}%
\bibitem [{\citenamefont {Wang}\ \emph {et~al.}()\citenamefont {Wang},
  \citenamefont {Edkins}, \citenamefont {Hamidian}, \citenamefont {Davis},
  \citenamefont {Fradkin},\ and\ \citenamefont {Kivelson}}]{Wang2018}%
  \BibitemOpen
\bibfield  {journal} {  }\bibfield  {author} {\bibinfo {author} {\bibfnamefont
  {Yuxuan}\ \bibnamefont {Wang}}, \bibinfo {author} {\bibfnamefont {Stephen}\
  \bibnamefont {Edkins}}, \bibinfo {author} {\bibfnamefont {Mohammad~H.}\
  \bibnamefont {Hamidian}}, \bibinfo {author} {\bibfnamefont {J.~C.~Seamus}\
  \bibnamefont {Davis}}, \bibinfo {author} {\bibfnamefont {Eduardo}\
  \bibnamefont {Fradkin}}, \ and\ \bibinfo {author} {\bibfnamefont {Steven~A.}\
  \bibnamefont {Kivelson}},\ }\href@noop {} {}\bibinfo {howpublished}
  {\url{arXiv:1204.1234}}\BibitemShut {NoStop}%
\end{thebibliography}%

\end{document}